\documentclass[12pt,preprint]{aa}

\usepackage{txfonts}
\usepackage{epsfig}
\usepackage{lscape}
\bibpunct{(}{)}{;}{a}{}{,}    
\usepackage{graphicx,graphics}
\usepackage{color}
\usepackage{ulem}
\usepackage{amsmath}
\usepackage{amssymb}
\usepackage[breaklinks=true, colorlinks=true, urlcolor=blue, citecolor=blue, linkcolor=blue]{hyperref}

\begin{document}

\title{Gas and stellar kinematic misalignment in MaNGA galaxies: what is the origin of counter-rotating gas?}

\author{
        I.~A.~Zinchenko\inst{\ref{LMU},\ref{MAO}}
       }
       
\institute{
Faculty of Physics, Ludwig-Maximilians-Universit\"{a}t, Scheinerstr. 1, 81679 Munich, Germany \label{LMU}
\and
Main Astronomical Observatory, National Academy of Sciences of Ukraine, 
27 Akademika Zabolotnoho St., 03143, Kyiv, Ukraine\label{MAO}
}

\abstract
   {Kinematic misalignment between gas and stellar components observed in a certain fraction of galaxies. It believed to be caused by acquisition of gas from the external reservoir by major or minor mergers, accretion from cosmological filaments or circumgalactic medium, etc. 
   }
   {We aim to constrain possible sources of the gas that forms counter-rotating component.}
   {We derived the gas-phase oxygen abundance in 69 galaxies with kinematic misalignment between gas and stellar components from MaNGA DR17 survey and compared it with the metallicity expected according to the mass-metallicity relation.}
   {We found that the oxygen abundance of the counter-rotating gas in our sample is higher than 8.2~dex that excludes significant role of inflow of pristine gas. Meanwhile, there is a significant difference in the oxygen abundance of the counter-rotating gas between red and blue galaxies. In general, the oxygen abundance is lower than expected for their stellar mass in red galaxies, but is compatible with or even higher than typical values for their stellar mass in blue galaxies.
   }
   {We showed that the exchange of enriched gas between galaxies is the most plausible mechanism for explaining the metallicity of counter-rotating gas components in galaxies of all masses and colors. Meanwhile, minor mergers may play a significant role in the formation of counter-rotating gas components in red and quenched galaxies.
   }


\titlerunning{...}
\authorrunning{Zinchenko et~al.}
\maketitle


\section{Introduction}

The study of galaxy kinematics has revealed the occurrence of counter-rotating components within certain galaxies. These components exhibit a rotational direction that is opposite to that of the galaxy's main body. This peculiar kinematic misalignment has been observed between stellar components \citep{Rubin1992, Rix1992, Bertola1996, Coccato2011, Johnston2013, Krajnovic2015} as well as between gaseous and stellar components \citep{Bettoni1984, Galletta1987, Ciri1995, Davis2011, Chung2012, Pizzella2018}.

A number of mechanisms have been suggested to explain the kinematic misalignment observed in galaxy components, starting with the hypothesis of purely internal redistribution of matter within the disk and bulge. These scenarios include resonance capturing \citep{Tremaine2000}, dissolving bars \citep{Evans1994}, and angular momentum exchange with a bar \citep{Pfenniger1991}. While these mechanisms offer potential explanations, they struggle to account for the significant differences in ages and metallicities observed in the stellar components \citep{Pizzella2014, Nedelchev2019}.

Another class of scenarios, which has been theoretically proposed to explain the formation of counter-rotating components, involves the acquisition of gas with retrograde rotation from regions outside the galactic disk. These scenarios include: i) major mergers of galaxies \citep{Puerari2001}; ii) gas-rich minor mergers \citep{Thakar1998, Bassett2017}; and iii) gas accretion from cosmological filaments or the gas-rich circumgalactic medium \citep{Algorry2014, Khoperskov2021}. These scenarios find support in a number of observations that favor either mergers with gas-rich galaxies \citep{Thakar1997, DiMatteo2008, Saburova2018}, or retrograde gas accretion \citep{Thakar1998, Chung2012, Nedelchev2019, Osman2017, Pizzella2018, Bevacqua2022}.

Despite these proposed scenarios, the origin of the gas forming counter-rotating components in galaxies and its relation to various processes such as gas accretion, interaction with other galaxies, stellar and AGN feedback, and other effects, remains unclear.

In this work, we aim to study the origin of the gaseous counter-rotation component at $z \sim 0$ using the MaNGA sample of galaxies. Specifically, we will investigate the chemical abundance of the counter-rotating gas, which may help to distinguish between pristine gas and enriched gas, providing valuable insights into the formation mechanisms involved.

The paper is structured as follows. In Section~\ref{sect:data}, we provide a detailed description of the data used and the criteria employed for sample selection. Section~\ref{sect:M-ur-OH} describes the stellar mass, color, and gas-phase metallicity of galaxies with counter-rotation. In Section~\ref{sect:discussion}, we discuss potential scenarios for the formation of counter-rotating gaseous components. Finally, Section~\ref{section:Summary} presents a summary of the main findings.

\section{Sample selection}
\label{sect:data}

In this study, we utilized a sample of galaxies obtained from the MaNGA SDSS DR17 survey \citep{SDSSDR17}. To derive chemical and kinematic properties we analysed the MaNGA spectra following prescriptions described in \citet{Zinchenko2016,Zinchenko2021}.
In brief, the stellar component in all spaxels is fitted using the public version of the STARLIGHT code \citep{CidFernandes2005,Mateus2006,Asari2007}
adapted for parallel processing of datacubes.
To fit the stellar spectra we used simple stellar population (SSP) spectra from the evolutionary synthesis models by \citet{BC03}. The best fit of the stellar spectrum has been used to obtain line-of-sight stellar velocity. Also, it has been subtracted from the observed spectrum to obtain a pure gas spectrum. 

To fit the emission lines we used our code ELF3D for emission line fitting in the optical spectra. The code is built upon the \textit{LMFIT} package \citep{LMFIT2014}, which offers a high-level interface for non-linear optimization and curve fitting problems. For each spectrum, we measured the fluxes of several emission lines, including 
[\ion{O}{II}]$\lambda\,\lambda$3727,3729,
H$\beta$,
[\ion{O}{III}]$\lambda$4959,
[\ion{O}{III}]$\lambda$5007,
[\ion{N}{II}]$\lambda$6548,
H$\alpha$,
[\ion{N}{II}]$\lambda$6584, and
[\ion{S}{II}]$\lambda$6717,6731. To account for interstellar reddening, we applied the analytical approximation of the Whitford interstellar reddening law \citep{Izotov1994}, assuming a Balmer line ratio of $\text{H}\alpha/\text{H}\beta = 2.86$. In cases where the measured value of $\text{H}\alpha/\text{H}\beta$ was less than 2.86, we set the reddening to zero.

Next, we used the H$\alpha$ emission line to construct the gas velocity field and compared it with the stellar velocity field obtained from the SSP fitting. Visual inspection of the velocity fields in all datacubes allowed us to select galaxies exhibiting a clear rotation pattern in both the stellar and gaseous components. Specifically, we selected galaxies where the gaseous and stellar components were either counter-rotating or exhibited significant kinematic misalignment (difference in position angle (PA) is around or greater than 90$^\circ$). 

To further refine our sample, we considered only galaxies that contain star-forming regions within the effective radius (R$_e$), enabling us to derive oxygen abundances using strong line methods. To identify spaxels associated with \ion{H}{II} regions, we applied the $\log$([\ion{O}{III}]$\lambda$5007/H$\beta$) -- $\log$([\ion{N}{II}]$\lambda$6584/H$\alpha$) diagram \citep{BPT} and utilized the dividing line proposed by \citet{Kauffmann2003}.

For the derivation of oxygen abundances, we selected only spectra with a signal-to-noise ratio $\text{S/N} > 4$ in all the [\ion{O}{II}]$\lambda,\lambda$3727,3729, H$\beta$, [\ion{O}{III}]$\lambda$5007, H$\alpha$, and [\ion{N}{II}]$\lambda$6584, [\ion{S}{II}]$\lambda$6717,6731 lines. We applied the empirical R calibration \citep{PilyuginGrebel2016} to obtain the oxygen abundance. Importantly, this calibration is effective across the entire metallicity range, including the low-metallicity regime expected in cases of gas accretion from cosmological filaments or low-metallicity dwarfs.
Given the potential variation of metallicity with galactocentric radius, we derived the oxygen abundance in two radial bins: R < 0.5 R$_e$ and 0.5 R$_e$ < R < R$_e$. However, the oxygen abundance measurements in both these bins are only available for a subset of galaxies, and we found no significant radial abundance gradient in those cases. Henceforth, we will refer to the average oxygen abundance within the R$_e$. We only considered galaxies with a minimum of 5 star-forming spaxels within this radius. Consequently, our final sample comprises 69 galaxies.

The stellar masses, effective radii, and colors of the MaNGA sample were obtained from the NASA-Sloan Atlas (NSA) catalog\footnote{\href{http://nsatlas.org}{nsatlas.org}}. We adopted NSA stellar masses derived by fitting the elliptical Petrosian fluxes using the K-correction method, assuming the \citet{Chabrier2003} initial mass function and utilizing simple stellar population models from \citet{BC03}. Adopted $R_e$ is defined as the Sersic 50\% light radius along the major axis in the $r$ band.
The inclination and position angle of the major axis of the galaxies were obtained from the Sersic fit to the surface brightness profile in the $r$ band.

\section{Stellar mass, color and metallicity}
\label{sect:M-ur-OH}

\begin{figure}
\resizebox{1.00\hsize}{!}{\includegraphics[angle=000]{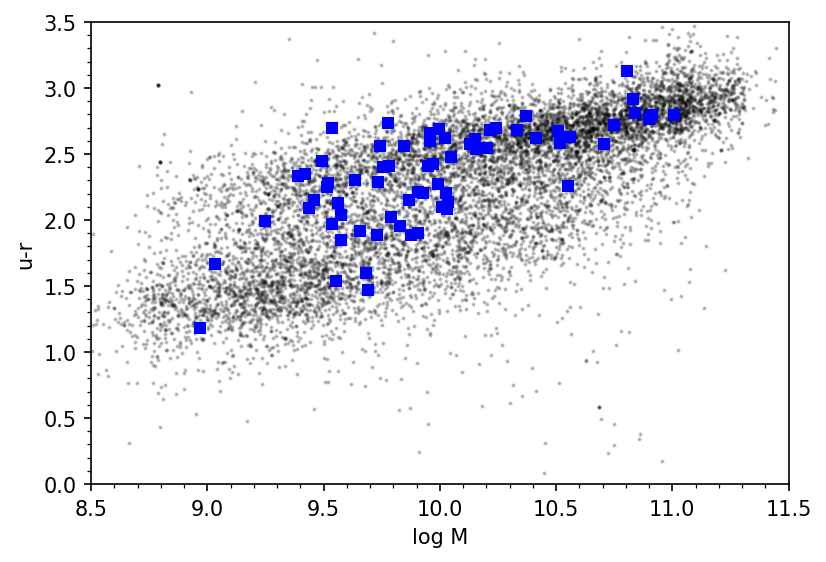}}
\caption{%
    The galaxy color - stellar mass diagram. Gray points represent the full sample of MaNGA galaxies. Blue squares are galaxies from our sample.
}
\label{figure:M-ur}
\end{figure}

In order to assess the distribution of our sample of galaxies with kinematic misalignment in terms of stellar mass and type, we employ the color-stellar mass diagram. As illustrated in Figure~\ref{figure:M-ur}, the majority of galaxies exhibit a high $u-r$ color, indicating a quenched or "red and dead" state. Only a small fraction of galaxies in our sample show signs of global and continuous star formation, as they reside in the blue cloud region and have low $u-r$ values. Notably, there is a relatively high fraction of galaxies from our sample located in the green valley, compared to the total number of MaNGA galaxies in that region. Additionally, it is worth mentioning that nearly all the galaxies with kinematic misalignment found in the green valley have a stellar mass of $\log(M/M_\odot) \sim 10.0$ or lower.

The predominance of red galaxies in our sample supports the hypothesis that the kinematically misaligned gas originated from outside the galaxy. However, this observation does not provide a conclusive distinction between merging events and the infall of extragalactic gas.

To further investigate the nature of the gas in the counter-rotating galaxies, we examine the stellar mass-metallicity diagram. It has been widely observed that galaxies follow a tight correlation on the stellar mass-metallicity diagram, with more massive galaxies generally exhibiting higher metallicities \citep{AndrewsMartini2013, Zinchenko2019, Zinchenko2021}. Therefore, the stellar mass of a galaxy with a counter-rotating gaseous disk reflects the host galaxy's properties, while the gas-phase metallicity reflects the history of the merging or accreting counter-rotating component. For instance, intergalactic gas within cosmological filaments is expected to retain its primordial chemical abundance. Therefore, galaxies that accreted gas from the cosmological filament or another galaxy with a different mass are expected to be outliers on the stellar mass - metallicity diagram. 

\begin{figure}
\resizebox{1.00\hsize}{!}{\includegraphics[angle=000]{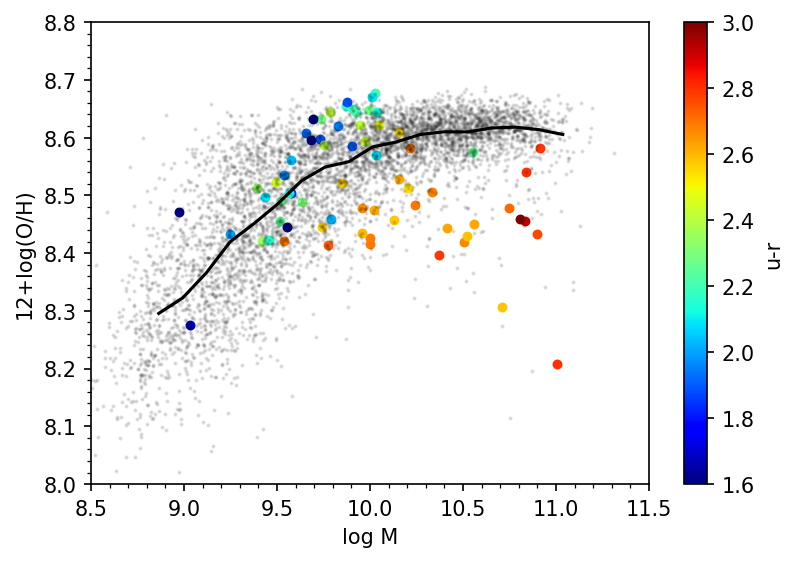}}
\caption{%
    The mass-metallicity diagram. Gray points represent the sample of MaNGA galaxies where oxygen abundance can be derived within R$_e$. Color circles are galaxies from our sample with $u-r$ showed in color. Solid black line represents median oxygen abundance in mass bins.
}
\label{figure:M-OH}
\end{figure}

In Figure~\ref{figure:M-OH}, we present the stellar mass-metallicity diagram for our sample of galaxies (color-coded symbols), in comparison to the overall MaNGA galaxy population (gray points). 
The $u-r$ color index is usually used in photometric studies to distinguish between quenched red galaxies and star-forming blue galaxies. Here we applied it to our sample of galaxies with kinematic misalignment between the gaseous and stellar components. Notably, none of the counter-rotating galaxies in our sample exhibit very low gas-phase metallicities. The lowest oxygen abundance of 8.21~dex is found in the most massive red galaxy with $\log(M/M_\odot) = 11.01$ and $u-r = 2.80$. The absence of low-metallicity gas suggests that the counter-rotating gas has undergone pre-processing through star formation cycles and is not pristine gas directly accreted from cosmological filaments.

Another intriguing feature seen in this plot is that not all galaxies in our sample deviate significantly from the average mass-metallicity relation. In fact, blue star-forming galaxies and those in the green valley ($u-r < 2.4$) exhibit similar or even higher gas-phase oxygen abundances compared to other galaxies of similar stellar masses. On the contrary, red galaxies with counter-rotating gaseous components demonstrate significantly lower gas-phase oxygen abundances relative to other galaxies of similar stellar masses.

Therefore, our sample of galaxies with kinematic misalignment between the gaseous and stellar components can be divided into two distinct subgroups with different behaviors on the mass-metallicity diagram. The primary distinction between these subgroups lies in their color and, consequently, their past star formation rates. This finding provides additional constraints on the potential mechanisms responsible for the formation of counter-rotating components in galaxies.

\section{Discussion}
\label{sect:discussion}

\subsection{The environment}

Based on the study of counter-rotating stellar disks in 64 galaxies from MaNGA DR16, \citet{Bevacqua2022} suggests that the formation of counter-rotating components in galaxies is likely influenced by the environment. Specifically, merger and interaction may be the main cause of counter-rotating components in galaxies in dense environments, while in low-density environments, undisturbed gas accretion, for example from cosmological filaments, may be more dominant in the formation of counter-rotating disks. Additionally, \citet{Beom2022} found that counter-rotating galaxies in MaNGA are preferentially found in small groups. 

According to the Galaxy Environment for MaNGA Value Added Catalog (GEMA-VAC), approximately 45\% of the galaxies in our sample belong to pairs or groups. However, this value may be a lower limit because visual inspection of the images of our sample of galaxies taken by SDSS and Legacy Surveys revealed the presence of tidal features or potential dwarf satellites for some galaxies classified as isolated in the GEMA-VAC catalog. We did not find a correlation between the oxygen abundance and membership in a pair or group, according to the GEMA-VAC catalog, except for the fact that the majority of massive galaxies are classified as members of pairs or groups. However, it is noteworthy that all massive galaxies in our sample exhibit lower oxygen abundances.

The significant number of galaxies in groups and/or exhibiting tidal features in our sample may suggest that the exchange of enriched gas between galaxies in groups, as observed between host galaxies and their satellites in \citet{Schaefer2019}, could also contribute to the formation of kinematic misalignment in the gas component. Overall, these results confirm that the environment may play an important role in formation of counter-rotating gas components.

\subsection{An example of interacting system}

\begin{figure}
\resizebox{1.00\hsize}{!}{\includegraphics[angle=000]{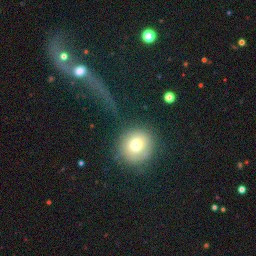}}
\caption{%
    Color composite image of VII~Zw~720 and SDSS~J173202.96+595854.7 made using three optical bands (g, r, z) taken from the Legacy Survey DR10 sky map.
}
\label{figure:image}
\end{figure}

Along with several galaxies in our sample exhibiting tidal features ($\sim$~25\%), we have identified one pair of galaxies that likely undergoing an exchange of baryonic matter. More massive galaxy in this pair, VII~Zw~720, is a quenched galaxy with a stellar mass of $\log(M/M_\sun) = 10.02$, $u-r = 2.61$ and a counter-rotating gas component. The other galaxy, SDSS~J173202.96+595854.7, is located 47 arcseconds to the northeast of VII~Zw~720 and exhibits a wide open spiral arm pointing towards VII~Zw~720, as illustrated in Figure~\ref{figure:image}. Fortunately, the center of SDSS~J173202.96+595854.7 has an available SDSS spectrum with a measured redshift $z = 0.02914$, which is remarkably close to the redshift of VII~Zw~720 ($z = 0.02920$). These facts suggest that these galaxies are most likely interacting and dynamically connected.

According to the NSA catalog, SDSS~J173202.96+595854.7 has a stellar mass of $\log(M/M_\sun) = 9.05$ and a $u-r = 1.68$. The SDSS spectrum obtained for its center enables us to derive the oxygen abundance in this galaxy. Analyzing the emission line ratios on the BPT diagram, we find that the spectrum falls within the composite region, although very close to the demarcation line between composite and Seyfert spectra proposed by \citet{Kewley2001}. Consequently, we adopt the oxygen abundance calibration for Seyferts developed by \citet{Dors2021}. The resulting oxygen abundances for both galaxies are remarkably similar, 8.49~dex for SDSS~J173202.96+595854.7 and 8.47~dex for VII~Zw~720. 

It should be noted that both galaxies appear as outliers on the mass-metallicity diagram. VII~Zw~720 has a higher stellar mass than expected at its metallicity, while SDSS~J173202.96+595854.7 has a lower stellar mass than anticipated based on the mass-metallicity relation. This suggests that the more massive and red galaxy, VII~Zw~720, may be accreting gas from its less massive but more gas-rich companion. During this process, the low mass galaxy could experience tidal disruption, leading to the loss of a fraction of its stellar mass that shifting the galaxy to an outlier region above the averaged mass-metallicity relation. Simultaneously, the acquisition of lower metallicity gas from the companion shifts VII~Zw 720 below the averaged mass-metallicity relation. Considering the current difference in stellar mass between these galaxies is approximately tenfold, this scenario could indicate the early stages of a minor merger event, leading to the formation of a counter-rotating gas component. Thus, minor mergers involving gas-rich galaxies may represent one of the primary mechanisms contributing to the formation of counter-rotating gas components, particularly in massive and quenched galaxies.

\subsection{Counter-rotating components of high metallicity}

Although the minor merger scenario provides a plausible explanation for the metallicity of the counter-rotating gas component in red galaxies, it becomes challenging to account for the origin of counter-rotating gas components in star-forming galaxies with low $u-r$ colors using minor mergers. In Figure~\ref{figure:M-OH} galaxies with low $u-r$ generally follow the averaged mass-metallicity relation or even exceed the average metallicity for their respective stellar masses.

In spite of the stellar components that are collisionless, a configuration with two counter-rotating gas components is unstable, and one of the gas components will quickly dissolve~\citep[see, e.g.,][]{Khoperskov2021}. Hence, for the formation of a counter-rotating gas component, the angular momentum of the merging gas component must exceed that of the co-rotating gas. Consequently, the mass and, therefore, metallicity of the merging gas should dominate in the newly formed counter-rotating gas component~\citep{Bassett2017}. This sets a lower limit on the amount of gas (and the minor-merger mass ratio) required to form a counter-rotating component in the presence of co-rotating gas in the host galaxy. However, defining an exact limit is challenging since even in the simpler case of a stellar merger without a gas component, the final configuration is dependent on orbit parameters and the mass ratio involved~\citep{Zinchenko2015}.

Additionally, the mass-metallicity relation implies that when a host galaxy merges with a gas-rich galaxy of lower mass, the resulting host galaxy will exhibit a lower metallicity. Therefore, in the case of a minor merger, the counter-rotating gas component, on average, should have a lower metallicity than expected for the host galaxy's mass. However, a major merger can produce counter-rotating gas components with metallicities consistent with the average mass-metallicity relation.

However, a major merger scenario should also produce two counter-rotating stellar components with comparable masses. To investigate this aspect within our sample of counter-rotating galaxies, we compared our sample with the one from \citet{Bevacqua2022}, who identified 64 galaxies with counter-rotating stellar disks among 4000 galaxies from MaNGA DR16. Interestingly, we found only five galaxies that appear in both samples, meaning they possess both counter-rotating gaseous and stellar components simultaneously.
Among these five galaxies, four exhibit low $u-r$ colors and oxygen abundances higher than the average value for their respective stellar masses. However, it remains unclear whether the counter-rotating stellar component was formed in-situ from counter-rotating gas or acquired from another galaxy during the merging process. Thus, while major mergers represent a potential channel for the formation of counter-rotating gaseous components in star-forming galaxies, they do not seem to be the dominant mechanism.

In addition to the mechanisms discussed earlier, the exchange of enriched gas between galaxies in groups or clusters, as suggested by \citet{Schaefer2019}, can also contribute to the formation of counter-rotating gas components. Figure~\ref{figure:M-OH} shows that the oxygen abundance of the counter-rotating gas component in galaxies within the intermediate stellar mass range of $\log(M/M_\sun) \sim 10.0$ varies from approximately 8.35 dex to 8.7 dex. Furthermore, there is a correlation between oxygen abundance and color, where red galaxies exhibit lower oxygen abundances compared to blue galaxies.
The exchange of enriched gas between galaxies can explain the existence of counter-rotating gas with both higher and lower metallicities compared to the average metallicity for a given galaxy mass, depending on the metallicity of the galaxy from which the gas originated. 

Quenched and red galaxies lacking a significant co-rotating gas disk may require the infall of a relatively small amount of counter-rotating gas to form a counter-rotating gas component. This gas could originate from the feedback of a low-mass galaxy with low metallicity. On the other hand, star-forming and blue galaxies already possess co-rotating gaseous disk and would require a larger amount of counter-rotating gas to reverse the rotation of the gas component. In such cases, the amount of gas ejected from a low-mass galaxy may be insufficient to replace the existing co-rotating gaseous disk. Instead, the formation of a counter-rotating gas component would necessitate the accretion of a significant amount of gas that can be provided by another galaxy of similar or higher mass. Particularly, high-metallicity galaxies in our sample with oxygen abundances above 8.6~dex in the mass range of $9.7 < \log(M/M_\sun) \sim 10.4$ may inherit their high-metallicity gas through feedback from more massive galaxies, for which such high metallicity is typical.

The transfer of enriched gas from one galaxy to another, where it may form a counter-rotating disk, can be facilitated by various physical mechanisms. These include stellar and AGN feedback~\citep{vandeVoort2016, Hafen2019, Choi2020, Li2023} as well as ram pressure stripping of gas in galaxy clusters~\citep{Gullieuszik2023, George2023}.

\section{Summary and conclusions}
\label{section:Summary}

We identified counter-rotating gas components containing \ion{H}{II}~regions in 69 galaxies from the MaNGA DR17 survey. To gain insights into the origin of these counter-rotating gas, we conducted a comparative analysis involving the stellar mass of the galaxy, its color index ($u-r$), and the oxygen abundance of the counter-rotating gas. Our analysis aimed to provide constraints on the origin of counter-rotating gas components. Our main conclusions are the following:

\begin{enumerate}

\item The oxygen abundance of the counter-rotating gas components in our sample is higher than 8.2~dex, suggesting that direct inflow of pristine gas from cosmological filaments is an unlikely source of gas acquisition for galaxies in the current epoch, especially for the formation of counter-rotating gas components.

\item The majority of galaxies with a counter-rotating gas component in our sample are either red and quenched, or lie in the green valley on the stellar mass--color diagram.

\item There is a significant difference in the oxygen abundance of the counter-rotating gas components hosted by red and blue galaxies. In general, the oxygen abundance is lower than expected for their stellar mass in red galaxies, but is compatible with or even higher than typical values for their stellar mass in blue galaxies.

\item Minor mergers may explain the low gas-phase metallicity of red galaxies, but cannot account for the higher metallicity of blue galaxies.

\item Major mergers and/or the exchange of enriched gas between galaxies are able to explain the oxygen abundance of blue galaxies. However, we did not find conclusive evidence for such disruptive events like major mergers in our sample.

\end{enumerate}

Therefore, the exchange of enriched gas between galaxies seems to be the most plausible mechanism for explaining the metallicity of counter-rotating gas components in galaxies of all masses and colors, while minor mergers may play a significant role in the formation of counter-rotating gas components in red and quenched galaxies.

\section*{Acknowledgements}

We are grateful to the referee for his/her constructive comments. \\
The authors gratefully acknowledge the computational and data resources provided by the Leibniz Supercomputing Centre (www.lrz.de). \\
SDSS-IV acknowledges support and resources from the Center for High-Performance Computing at the University of Utah.
The SDSS web site is \href{https://www.sdss.org/}{www.sdss.org}. \\
SDSS-IV is managed by the Astrophysical Research Consortium for the 
Participating Institutions of the SDSS Collaboration including the 
Brazilian Participation Group, the Carnegie Institution for Science, Carnegie Mellon University, the Chilean Participation Group, the French Participation Group, Harvard-Smithsonian Center for Astrophysics, Instituto de Astrof\'{\i}sica de Canarias, The Johns Hopkins University, Kavli Institute for the Physics and Mathematics of the Universe (IPMU) / 
University of Tokyo, Lawrence Berkeley National Laboratory, 
Leibniz Institut f\"ur Astrophysik Potsdam (AIP),  
Max-Planck-Institut f\"ur Astronomie (MPIA Heidelberg), 
Max-Planck-Institut f\"ur Astrophysik (MPA Garching), 
Max-Planck-Institut f\"ur Extraterrestrische Physik (MPE), 
National Astronomical Observatories of China, New Mexico State University, New York University, University of Notre Dame, Observat\'orio Nacional / MCTI, The Ohio State University, Pennsylvania State University, Shanghai Astronomical Observatory, United Kingdom Participation Group, Universidad Nacional Aut\'onoma de M\'exico, University of Arizona, University of Colorado Boulder, University of Oxford, University of Portsmouth, University of Utah, University of Virginia, University of Washington, University of Wisconsin, Vanderbilt University, and Yale University. \\

\bibliography{reference}

\begin{thebibliography}{53}
\expandafter\ifx\csname natexlab\endcsname\relax\def\natexlab#1{#1}\fi

\bibitem[{{Abdurro'uf} {et~al.}(2022){Abdurro'uf}, {Accetta}, {Aerts}, {Silva
  Aguirre}, {Ahumada}, {Ajgaonkar}, {Filiz Ak}, {Alam}, {Allende Prieto},
  {Almeida}, {Anders}, {Anderson}, {Andrews}, {Anguiano}, {Aquino-Ort{\'\i}z},
  {Arag{\'o}n-Salamanca}, {Argudo-Fern{\'a}ndez}, {Ata}, {Aubert},
  {Avila-Reese}, {Badenes}, {Barb{\'a}}, {Barger}, {Barrera-Ballesteros},
  {Beaton}, {Beers}, {Belfiore}, {Bender}, {Bernardi}, {Bershady}, {Beutler},
  {Bidin}, {Bird}, {Bizyaev}, {Blanc}, {Blanton}, {Boardman}, {Bolton},
  {Boquien}, {Borissova}, {Bovy}, {Brandt}, {Brown}, {Brownstein}, {Brusa},
  {Buchner}, {Bundy}, {Burchett}, {Bureau}, {Burgasser}, {Cabang}, {Campbell},
  {Cappellari}, {Carlberg}, {Wanderley}, {Carrera}, {Cash}, {Chen}, {Chen},
  {Cherinka}, {Chiappini}, {Choi}, {Chojnowski}, {Chung}, {Clerc}, {Cohen},
  {Comerford}, {Comparat}, {da Costa}, {Covey}, {Crane}, {Cruz-Gonzalez},
  {Culhane}, {Cunha}, {Dai}, {Damke}, {Darling}, {Davidson}, {Davies},
  {Dawson}, {De Lee}, {Diamond-Stanic}, {Cano-D{\'\i}az}, {S{\'a}nchez},
  {Donor}, {Duckworth}, {Dwelly}, {Eisenstein}, {Elsworth}, {Emsellem},
  {Eracleous}, {Escoffier}, {Fan}, {Farr}, {Feng}, {Fern{\'a}ndez-Trincado},
  {Feuillet}, {Filipp}, {Fillingham}, {Frinchaboy}, {Fromenteau}, {Galbany},
  {Garc{\'\i}a}, {Garc{\'\i}a-Hern{\'a}ndez}, {Ge}, {Geisler}, {Gelfand},
  {G{\'e}ron}, {Gibson}, {Goddy}, {Godoy-Rivera}, {Grabowski}, {Green},
  {Greener}, {Grier}, {Griffith}, {Guo}, {Guy}, {Hadjara}, {Harding},
  {Hasselquist}, {Hayes}, {Hearty}, {Hern{\'a}ndez}, {Hill}, {Hogg},
  {Holtzman}, {Horta}, {Hsieh}, {Hsu}, {Hsu}, {Huber}, {Huertas-Company},
  {Hutchinson}, {Hwang}, {Ibarra-Medel}, {Chitham}, {Ilha}, {Imig}, {Jaekle},
  {Jayasinghe}, {Ji}, {Johnson}, {Jones}, {J{\"o}nsson}, {Katkov}, {Khalatyan},
  {Kinemuchi}, {Kisku}, {Knapen}, {Kneib}, {Kollmeier}, {Kong}, {Kounkel},
  {Kreckel}, {Krishnarao}, {Lacerna}, {Lane}, {Langgin}, {Lavender}, {Law},
  {Lazarz}, {Leung}, {Leung}, {Lewis}, {Li}, {Li}, {Lian}, {Liang}, {Lin},
  {Lin}, {Lin}, {Lintott}, {Long}, {Longa-Pe{\~n}a}, {L{\'o}pez-Cob{\'a}},
  {Lu}, {Lundgren}, {Luo}, {Mackereth}, {de la Macorra}, {Mahadevan},
  {Majewski}, {Manchado}, {Mandeville}, {Maraston}, {Margalef-Bentabol},
  {Masseron}, {Masters}, {Mathur}, {McDermid}, {Mckay}, {Merloni},
  {Merrifield}, {Meszaros}, {Miglio}, {Di Mille}, {Minniti}, {Minsley},
  {Monachesi}, {Moon}, {Mosser}, {Mulchaey}, {Muna}, {Mu{\~n}oz}, {Myers},
  {Myers}, {Nadathur}, {Nair}, {Nandra}, {Neumann}, {Newman}, {Nidever},
  {Nikakhtar}, {Nitschelm}, {O'Connell}, {Garma-Oehmichen}, {Luan Souza de
  Oliveira}, {Olney}, {Oravetz}, {Ortigoza-Urdaneta}, {Osorio}, {Otter},
  {Pace}, {Padilla}, {Pan}, {Pan}, {Parikh}, {Parker}, {Peirani}, {Pe{\~n}a
  Ram{\'\i}rez}, {Penny}, {Percival}, {Perez-Fournon}, {Pinsonneault},
  {Poidevin}, {Poovelil}, {Price-Whelan}, {B{\'a}rbara de Andrade Queiroz},
  {Raddick}, {Ray}, {Rembold}, {Riddle}, {Riffel}, {Riffel}, {Rix}, {Robin},
  {Rodr{\'\i}guez-Puebla}, {Roman-Lopes}, {Rom{\'a}n-Z{\'u}{\~n}iga}, {Rose},
  {Ross}, {Rossi}, {Rubin}, {Salvato}, {S{\'a}nchez}, {S{\'a}nchez-Gallego},
  {Sanderson}, {Santana Rojas}, {Sarceno}, {Sarmiento}, {Sayres}, {Sazonova},
  {Schaefer}, {Schiavon}, {Schlegel}, {Schneider}, {Schultheis}, {Schwope},
  {Serenelli}, {Serna}, {Shao}, {Shapiro}, {Sharma}, {Shen}, {Shetrone}, {Shu},
  {Simon}, {Skrutskie}, {Smethurst}, {Smith}, {Sobeck}, {Spoo}, {Sprague},
  {Stark}, {Stassun}, {Steinmetz}, {Stello}, {Stone-Martinez},
  {Storchi-Bergmann}, {Stringfellow}, {Stutz}, {Su}, {Taghizadeh-Popp},
  {Talbot}, {Tayar}, {Telles}, {Teske}, {Thakar}, {Theissen}, {Tkachenko},
  {Thomas}, {Tojeiro}, {Hernandez Toledo}, {Troup}, {Trump}, {Trussler},
  {Turner}, {Tuttle}, {Unda-Sanzana}, {V{\'a}zquez-Mata}, {Valentini},
  {Valenzuela}, {Vargas-Gonz{\'a}lez}, {Vargas-Maga{\~n}a}, {Alfaro},
  {Villanova}, {Vincenzo}, {Wake}, {Warfield}, {Washington}, {Weaver},
  {Weijmans}, {Weinberg}, {Weiss}, {Westfall}, {Wild}, {Wilde}, {Wilson},
  {Wilson}, {Wilson}, {Wolf}, {Wood-Vasey}, {Yan}, {Zamora}, {Zasowski},
  {Zhang}, {Zhao}, {Zheng}, {Zheng}, \& {Zhu}}]{SDSSDR17}
{Abdurro'uf}, {Accetta}, K., {Aerts}, C., {et~al.} 2022, \apjs, 259, 35

\bibitem[{{Algorry} {et~al.}(2014){Algorry}, {Navarro}, {Abadi}, {Sales},
  {Steinmetz}, \& {Piontek}}]{Algorry2014}
{Algorry}, D.~G., {Navarro}, J.~F., {Abadi}, M.~G., {et~al.} 2014, \mnras, 437,
  3596

\bibitem[{{Andrews} \& {Martini}(2013)}]{AndrewsMartini2013}
{Andrews}, B.~H. \& {Martini}, P. 2013, \apj, 765, 140

\bibitem[{{Asari} {et~al.}(2007){Asari}, {Cid Fernandes}, {Stasi{\'n}ska},
  {Torres-Papaqui}, {Mateus}, {Sodr{\'e}}, {Schoenell}, \& {Gomes}}]{Asari2007}
{Asari}, N.~V., {Cid Fernandes}, R., {Stasi{\'n}ska}, G., {et~al.} 2007,
  \mnras, 381, 263

\bibitem[{{Baldwin} {et~al.}(1981){Baldwin}, {Phillips}, \& {Terlevich}}]{BPT}
{Baldwin}, J.~A., {Phillips}, M.~M., \& {Terlevich}, R. 1981, \pasp, 93, 5

\bibitem[{{Bassett} {et~al.}(2017){Bassett}, {Bekki}, {Cortese}, \&
  {Couch}}]{Bassett2017}
{Bassett}, R., {Bekki}, K., {Cortese}, L., \& {Couch}, W. 2017, \mnras, 471,
  1892

\bibitem[{{Beom} {et~al.}(2022){Beom}, {Bizyaev}, {Walterbos}, \&
  {Chen}}]{Beom2022}
{Beom}, M., {Bizyaev}, D., {Walterbos}, R. A.~M., \& {Chen}, Y. 2022, \mnras,
  516, 3175

\bibitem[{{Bertola} {et~al.}(1996){Bertola}, {Cinzano}, {Corsini}, {Pizzella},
  {Persic}, \& {Salucci}}]{Bertola1996}
{Bertola}, F., {Cinzano}, P., {Corsini}, E.~M., {et~al.} 1996, \apjl, 458, L67

\bibitem[{{Bettoni}(1984)}]{Bettoni1984}
{Bettoni}, D. 1984, The Messenger, 37, 17

\bibitem[{{Bevacqua} {et~al.}(2022){Bevacqua}, {Cappellari}, \&
  {Pellegrini}}]{Bevacqua2022}
{Bevacqua}, D., {Cappellari}, M., \& {Pellegrini}, S. 2022, \mnras, 511, 139

\bibitem[{{Bruzual} \& {Charlot}(2003)}]{BC03}
{Bruzual}, G. \& {Charlot}, S. 2003, \mnras, 344, 1000

\bibitem[{{Chabrier}(2003)}]{Chabrier2003}
{Chabrier}, G. 2003, \pasp, 115, 763

\bibitem[{{Choi} {et~al.}(2020){Choi}, {Brennan}, {Somerville}, {Ostriker},
  {Hirschmann}, \& {Naab}}]{Choi2020}
{Choi}, E., {Brennan}, R., {Somerville}, R.~S., {et~al.} 2020, \apj, 904, 8

\bibitem[{{Chung} {et~al.}(2012){Chung}, {Bureau}, {van Gorkom}, \&
  {Koribalski}}]{Chung2012}
{Chung}, A., {Bureau}, M., {van Gorkom}, J.~H., \& {Koribalski}, B. 2012,
  \mnras, 422, 1083

\bibitem[{{Cid Fernandes} {et~al.}(2005){Cid Fernandes}, {Mateus}, {Sodr{\'e}},
  {Stasi{\'n}ska}, \& {Gomes}}]{CidFernandes2005}
{Cid Fernandes}, R., {Mateus}, A., {Sodr{\'e}}, L., {Stasi{\'n}ska}, G., \&
  {Gomes}, J.~M. 2005, \mnras, 358, 363

\bibitem[{{Ciri} {et~al.}(1995){Ciri}, {Bettoni}, \& {Galletta}}]{Ciri1995}
{Ciri}, R., {Bettoni}, D., \& {Galletta}, G. 1995, \nat, 375, 661

\bibitem[{{Coccato} {et~al.}(2011){Coccato}, {Morelli}, {Corsini}, {Buson},
  {Pizzella}, {Vergani}, \& {Bertola}}]{Coccato2011}
{Coccato}, L., {Morelli}, L., {Corsini}, E.~M., {et~al.} 2011, \mnras, 412,
  L113

\bibitem[{{Davis} {et~al.}(2011){Davis}, {Alatalo}, {Sarzi}, {Bureau}, {Young},
  {Blitz}, {Serra}, {Crocker}, {Krajnovi{\'c}}, {McDermid}, {Bois}, {Bournaud},
  {Cappellari}, {Davies}, {Duc}, {de Zeeuw}, {Emsellem}, {Khochfar},
  {Kuntschner}, {Lablanche}, {Morganti}, {Naab}, {Oosterloo}, {Scott}, \&
  {Weijmans}}]{Davis2011}
{Davis}, T.~A., {Alatalo}, K., {Sarzi}, M., {et~al.} 2011, \mnras, 417, 882

\bibitem[{{Di Matteo} {et~al.}(2008){Di Matteo}, {Combes}, {Melchior}, \&
  {Semelin}}]{DiMatteo2008}
{Di Matteo}, P., {Combes}, F., {Melchior}, A.~L., \& {Semelin}, B. 2008, \aap,
  477, 437

\bibitem[{{Dors}(2021)}]{Dors2021}
{Dors}, O.~L. 2021, \mnras, 507, 466

\bibitem[{{Evans} \& {Collett}(1994)}]{Evans1994}
{Evans}, N.~W. \& {Collett}, J.~L. 1994, \apjl, 420, L67

\bibitem[{{Galletta}(1987)}]{Galletta1987}
{Galletta}, G. 1987, \apj, 318, 531

\bibitem[{{George} {et~al.}(2023){George}, {Poggianti}, {Tomi{\v{c}}i{\'c}},
  {Postma}, {C{\^o}t{\'e}}, {Fritz}, {Ghosh}, {Gullieuszik}, {Hutchings},
  {Moretti}, {Omizzolo}, {Radovich}, {Sreekumar}, {Subramaniam}, {Tandon}, \&
  {Vulcani}}]{George2023}
{George}, K., {Poggianti}, B.~M., {Tomi{\v{c}}i{\'c}}, N., {et~al.} 2023,
  \mnras, 519, 2426

\bibitem[{{Gullieuszik} {et~al.}(2023){Gullieuszik}, {Giunchi}, {Poggianti},
  {Moretti}, {Scarlata}, {Calzetti}, {Werle}, {Zanella}, {Radovich},
  {Bellhouse}, {Bettoni}, {Franchetto}, {Fritz}, {Jaff{\'e}}, {McGee},
  {Mingozzi}, {Omizzolo}, {Tonnesen}, {Verheijen}, \&
  {Vulcani}}]{Gullieuszik2023}
{Gullieuszik}, M., {Giunchi}, E., {Poggianti}, B.~M., {et~al.} 2023, \apj, 945,
  54

\bibitem[{{Hafen} {et~al.}(2019){Hafen}, {Faucher-Gigu{\`e}re},
  {Angl{\'e}s-Alc{\'a}zar}, {Stern}, {Kere{\v{s}}}, {Hummels}, {Esmerian},
  {Garrison-Kimmel}, {El-Badry}, {Wetzel}, {Chan}, {Hopkins}, \&
  {Murray}}]{Hafen2019}
{Hafen}, Z., {Faucher-Gigu{\`e}re}, C.-A., {Angl{\'e}s-Alc{\'a}zar}, D.,
  {et~al.} 2019, \mnras, 488, 1248

\bibitem[{{Izotov} {et~al.}(1994){Izotov}, {Thuan}, \&
  {Lipovetsky}}]{Izotov1994}
{Izotov}, Y.~I., {Thuan}, T.~X., \& {Lipovetsky}, V.~A. 1994, \apj, 435, 647

\bibitem[{{Johnston} {et~al.}(2013){Johnston}, {Merrifield},
  {Arag{\'o}n-Salamanca}, \& {Cappellari}}]{Johnston2013}
{Johnston}, E.~J., {Merrifield}, M.~R., {Arag{\'o}n-Salamanca}, A., \&
  {Cappellari}, M. 2013, \mnras, 428, 1296

\bibitem[{{Kauffmann} {et~al.}(2003){Kauffmann}, {Heckman}, {Tremonti},
  {Brinchmann}, {Charlot}, {White}, {Ridgway}, {Brinkmann}, {Fukugita}, {Hall},
  {Ivezi{\'c}}, {Richards}, \& {Schneider}}]{Kauffmann2003}
{Kauffmann}, G., {Heckman}, T.~M., {Tremonti}, C., {et~al.} 2003, \mnras, 346,
  1055

\bibitem[{{Kewley} {et~al.}(2001){Kewley}, {Dopita}, {Sutherland}, {Heisler},
  \& {Trevena}}]{Kewley2001}
{Kewley}, L.~J., {Dopita}, M.~A., {Sutherland}, R.~S., {Heisler}, C.~A., \&
  {Trevena}, J. 2001, \apj, 556, 121

\bibitem[{{Khoperskov} {et~al.}(2021){Khoperskov}, {Zinchenko}, {Avramov},
  {Khrapov}, {Berczik}, {Saburova}, {Ishchenko}, {Khoperskov}, {Pulsoni},
  {Venichenko}, {Bizyaev}, \& {Moiseev}}]{Khoperskov2021}
{Khoperskov}, S., {Zinchenko}, I., {Avramov}, B., {et~al.} 2021, \mnras, 500,
  3870

\bibitem[{{Krajnovi{\'c}} {et~al.}(2015){Krajnovi{\'c}}, {Weilbacher},
  {Urrutia}, {Emsellem}, {Carollo}, {Shirazi}, {Bacon}, {Contini}, {Epinat},
  {Kamann}, {Martinsson}, \& {Steinmetz}}]{Krajnovic2015}
{Krajnovi{\'c}}, D., {Weilbacher}, P.~M., {Urrutia}, T., {et~al.} 2015, \mnras,
  452, 2

\bibitem[{{Li} {et~al.}(2023){Li}, {Fraternali}, {Marasco}, {Trager},
  {Pezzulli}, {Mancera Pi{\~n}a}, \& {Verheijen}}]{Li2023}
{Li}, A., {Fraternali}, F., {Marasco}, A., {et~al.} 2023, \mnras, 520, 147

\bibitem[{{Mateus} {et~al.}(2006){Mateus}, {Sodr{\'e}}, {Cid Fernandes},
  {Stasi{\'n}ska}, {Schoenell}, \& {Gomes}}]{Mateus2006}
{Mateus}, A., {Sodr{\'e}}, L., {Cid Fernandes}, R., {et~al.} 2006, \mnras, 370,
  721

\bibitem[{{Nedelchev} {et~al.}(2019){Nedelchev}, {Coccato}, {Corsini}, {Sarzi},
  {de Zeeuw}, {Pizzella}, {Dalla Bont{\`a}}, {Iodice}, \&
  {Morelli}}]{Nedelchev2019}
{Nedelchev}, B., {Coccato}, L., {Corsini}, E.~M., {et~al.} 2019, \aap, 623, A87

\bibitem[{{Newville} {et~al.}(2014){Newville}, {Stensitzki}, {Allen}, \&
  {Ingargiola}}]{LMFIT2014}
{Newville}, M., {Stensitzki}, T., {Allen}, D.~B., \& {Ingargiola}, A. 2014,
  {LMFIT: Non-Linear Least-Square Minimization and Curve-Fitting for Python},
  Zenodo

\bibitem[{{Osman} \& {Bekki}(2017)}]{Osman2017}
{Osman}, O. \& {Bekki}, K. 2017, \mnras, 471, L87

\bibitem[{{Pfenniger} \& {Friedli}(1991)}]{Pfenniger1991}
{Pfenniger}, D. \& {Friedli}, D. 1991, \aap, 252, 75

\bibitem[{{Pilyugin} \& {Grebel}(2016)}]{PilyuginGrebel2016}
{Pilyugin}, L.~S. \& {Grebel}, E.~K. 2016, \mnras, 457, 3678

\bibitem[{{Pizzella} {et~al.}(2018){Pizzella}, {Morelli}, {Coccato}, {Corsini},
  {Dalla Bont{\`a}}, {Fabricius}, \& {Saglia}}]{Pizzella2018}
{Pizzella}, A., {Morelli}, L., {Coccato}, L., {et~al.} 2018, \aap, 616, A22

\bibitem[{{Pizzella} {et~al.}(2014){Pizzella}, {Morelli}, {Corsini}, {Dalla
  Bont{\`a}}, {Coccato}, \& {Sanjana}}]{Pizzella2014}
{Pizzella}, A., {Morelli}, L., {Corsini}, E.~M., {et~al.} 2014, \aap, 570, A79

\bibitem[{{Puerari} \& {Pfenniger}(2001)}]{Puerari2001}
{Puerari}, I. \& {Pfenniger}, D. 2001, \apss, 276, 909

\bibitem[{{Rix} {et~al.}(1992){Rix}, {Franx}, {Fisher}, \&
  {Illingworth}}]{Rix1992}
{Rix}, H.-W., {Franx}, M., {Fisher}, D., \& {Illingworth}, G. 1992, \apjl, 400,
  L5

\bibitem[{{Rubin} {et~al.}(1992){Rubin}, {Graham}, \& {Kenney}}]{Rubin1992}
{Rubin}, V.~C., {Graham}, J.~A., \& {Kenney}, J. D.~P. 1992, \apjl, 394, L9

\bibitem[{{Saburova} {et~al.}(2018){Saburova}, {Chilingarian}, {Katkov},
  {Egorov}, {Kasparova}, {Khoperskov}, {Uklein}, \& {Vozyakova}}]{Saburova2018}
{Saburova}, A.~S., {Chilingarian}, I.~V., {Katkov}, I.~Y., {et~al.} 2018,
  \mnras, 481, 3534

\bibitem[{{Schaefer} {et~al.}(2019){Schaefer}, {Tremonti}, {Pace}, {Belfiore},
  {Argudo-Fernandez}, {Bershady}, {Drory}, {Jones}, {Maiolino}, {Stark},
  {Wake}, \& {Yan}}]{Schaefer2019}
{Schaefer}, A.~L., {Tremonti}, C., {Pace}, Z., {et~al.} 2019, \apj, 884, 156

\bibitem[{{Thakar} \& {Ryden}(1998)}]{Thakar1998}
{Thakar}, A.~R. \& {Ryden}, B.~S. 1998, \apj, 506, 93

\bibitem[{{Thakar} {et~al.}(1997){Thakar}, {Ryden}, {Jore}, \&
  {Broeils}}]{Thakar1997}
{Thakar}, A.~R., {Ryden}, B.~S., {Jore}, K.~P., \& {Broeils}, A.~H. 1997, \apj,
  479, 702

\bibitem[{{Tremaine} \& {Yu}(2000)}]{Tremaine2000}
{Tremaine}, S. \& {Yu}, Q. 2000, \mnras, 319, 1

\bibitem[{{van de Voort} {et~al.}(2016){van de Voort}, {Quataert}, {Hopkins},
  {Faucher-Gigu{\`e}re}, {Feldmann}, {Kere{\v{s}}}, {Chan}, \&
  {Hafen}}]{vandeVoort2016}
{van de Voort}, F., {Quataert}, E., {Hopkins}, P.~F., {et~al.} 2016, \mnras,
  463, 4533

\bibitem[{{Zinchenko} {et~al.}(2015){Zinchenko}, {Berczik}, {Grebel},
  {Pilyugin}, \& {Just}}]{Zinchenko2015}
{Zinchenko}, I.~A., {Berczik}, P., {Grebel}, E.~K., {Pilyugin}, L.~S., \&
  {Just}, A. 2015, \apj, 806, 267

\bibitem[{{Zinchenko} {et~al.}(2019){Zinchenko}, {Just}, {Pilyugin}, \&
  {Lara-Lopez}}]{Zinchenko2019}
{Zinchenko}, I.~A., {Just}, A., {Pilyugin}, L.~S., \& {Lara-Lopez}, M.~A. 2019,
  \aap, 623, A7

\bibitem[{{Zinchenko} {et~al.}(2016){Zinchenko}, {Pilyugin}, {Grebel},
  {S{\'a}nchez}, \& {V{\'{\i}}lchez}}]{Zinchenko2016}
{Zinchenko}, I.~A., {Pilyugin}, L.~S., {Grebel}, E.~K., {S{\'a}nchez}, S.~F.,
  \& {V{\'{\i}}lchez}, J.~M. 2016, \mnras, 462, 2715

\bibitem[{{Zinchenko} {et~al.}(2021){Zinchenko}, {V{\'\i}lchez},
  {P{\'e}rez-Montero}, {Sukhorukov}, {Sobolenko}, \& {Duarte
  Puertas}}]{Zinchenko2021}
{Zinchenko}, I.~A., {V{\'\i}lchez}, J.~M., {P{\'e}rez-Montero}, E., {et~al.}
  2021, \aap, 655, A58

\end{thebibliography}
 
\end{document}